\documentclass[a4paper]{jpconf}
\usepackage{graphicx}
\begin{document}
\title{Direct numerical computation and its application to the higher-order radiative corrections}

\author{K Kato$^{1}$, E de Doncker$^{2}$, T Ishikawa$^{3}$  and F Yuasa$^{3}$}

\address{$^{1}$ Kogakuin University, Shinjuku, Tokyo 163-8677, Japan}
\address{$^{2}$ Western Michigan University,%
 1903 West Michigan Avenue, Kalamazoo, MI 49008, United States}
\address{$^{3}$ KEK, Tsukuba, Ibaraki 305-0801, Japan}

\ead{kato@cc.kogakuin.ac.jp}

\begin{abstract}
The direct computation method(DCM) is developed to calculate 
the multi-loop amplitude for general masses and external
momenta. The ultraviolet divergence is under control in
dimensional regularization.
In this paper we report on the progress of DCM to several scalar multi-loop integrals
after the presentation in ACAT2016.
Also the discussion is given on the application of DCM to
physical 2-loop processes including numerator functions.
\end{abstract}

\newcommand{\siki}[1]{Eq.(\ref{eq:#1})}
\newcommand{\zu}[1]{Fig.\ref{fig:#1}}
\newcommand{\hyo}[1]{Tab.\ref{tab:#1}}
\newcommand{\setu}[1]{Sec.\ref{sec:#1}}

\section{Introduction}
For the investigation of physics 
in the current and future collider experiments,
a precise evaluation of higher order corrections in perturbative quantum field theory(QFT) is required.
The calculation of the higher order radiative corrections
turns out to be a large scale computation since
the required number of Feynman diagrams is huge and the integral
of each diagram is sometimes very complicated.
The role of an automated system for the calculation
of perturbative series in QFT is not only to manage such a large
scale computation but to avoid possible errors caused by humans.

The starting point of the study in high-energy physics is QFT, i.e.,
the symbolic representation of the theory and the final output is
the predicted numerical values to be compared with experimental results.
So the specific feature of the system is depicted at which point one
switches from the symbolic treatment to the numerical one.
In this sense, our system would be described as the maximally
numerical method.

The library of the multi-loop integrals is an important
component of the automated system and the libraries are
working well in the 1-loop calculations\cite{GRACE,FeynArt,FDC,vanHameren, Golem95C, Collier, QCDLoop}.
Beyond 1-loop,
we have worked on the development of a computational method for Feynman loop integrals 
with a fully numerical approach. 
It is based on numerical integration and extrapolation techniques. In this paper, 
we describe the status and new developments
in our techniques for the numerical computation 
of Feynman loop integrals.

\section{Direct computation method}
The multi-loop integral is essential 
for the higher-order radiative corrections.
The integral can have singularities originating from the physics.

\begin{itemize}

\item[$-$] The function in the denominator can vanish, which is normally
avoided by the analytic continuation with $m \rightarrow m - i0$ in the analytical method.

\item[$-$] The ultraviolet(UV) divergence can appear when the integral
is divergent for the large momentum region when it is
calculated in 4 space-time dimension. 

\item[$-$] The infrared(IR) divergence can appear when a massless particle is
included in the integral and the integral is divergent in the soft momentum region.

\end{itemize}

\begin{table}
\caption{\label{tab:lists} The progress of DCM at ACAT2017.  %
The number shown as '$x$-dim.' stands for the maximum dimension of integrals.}
\begin{center}
\begin{tabular}{|c|p{41mm}|p{41mm}|p{41mm}|}
\hline
 & 2-point(self-energy) & 3-point(vertex) & 4-point(box) \\ \hline
       & massless, massive            &  massive   & massive   \\
2-loop & ref.\cite{twoSEms,ACAT2014}  & ref.\cite{twoVXms1,twoVXms2,ACAT2016}  & ref.\cite{twoBXms,twoBXms2}  \\
       & $4$-dim.                        & $5$-dim. & $6$-dim.  \\
\hline
       & massless, massive  &  massless \\
3-loop & ref.\cite{ACAT2016,threeSEml}  & ref.\cite{threeVXml,threeVXml2} \\
       & $7$-dim.  &  $6$-dim.  \\
\cline{1-3}
       & massless   \\
4-loop & ref.\cite{ACAT2016}  \\
       & $8$-dim.   \\
\cline{1-2}
\end{tabular}
\end{center}
\end{table}

The singularity in the integrand disappears if we introduce the regularization,
e.g., $m^2\rightarrow m^2-i\rho$, taking spacetime dimension to be $4-2\varepsilon$, 
or the introduction of the fictitious mass $\lambda$ for a massless particle.
With non-zero $\rho$,  $\varepsilon$ and $\lambda$ the integral
can be computed numerically and the physical value can be obtained by
extrapolation to the limit $\rho\rightarrow 0$ and so forth.

It is already shown that DCM can handle these singularities numerically.
For the scalar integrals the status is shown in Table.1.
After the last ACAT\cite{ACAT2016}, we have filled the box for
3-loop vertex functions.
The complexity of numerical integration increases with the dimension
of the integral and we have computed integrals up to dimension 8, or
a diagram with 9 propagators.
For the numerical integration, we use robust integration software in ParInt\cite{ParInt}
or the double exponential transformation method\cite{DE} and use
MPI\cite{openmpi} or other parallel environments for  accelerating the computation.

An important feature of DCM is that one does not need to separate
terms by hand as is done in the analytic treatment.
In a large-scale computation, manual operation is the point
where some error can happen.  
Suppose an integral $I$ has a UV singularity as
\begin{equation}
I=   \frac{C_{-K}}{\varepsilon^K} + \cdots  + \frac{C_{-1}}{\varepsilon} + C_0
 + C_1 \varepsilon + C_2 \varepsilon^2 + \cdots
\end{equation}
where we take the spacetime
dimension to be $n=4-2\varepsilon$.
Then, we compute $I$ as the integral whose integrand includes the numerical value of $\varepsilon$.
From a set of numerical values, we can obtain all values of the leading coefficients  
using a linear solver or Wynn's algorithm\cite{Wynn}.
When the most singular term is $1/\varepsilon^K$,
in the first case, we solve the linear equation
\begin{equation}
I(\varepsilon_j)=\sum_{k=-K}^{-K+N-1} C_k \varepsilon^k \qquad (j=1,\ldots,N)
\end{equation}
using an appropriate linear solver such as  {\tt dgefs.f} from the SLATEC 
Common Mathematical Library\cite{slatec}, 
and in the second case the following iteration is performed
\begin{equation}
a(j,k+1)=a(j+1,k-1)+\frac{1}{a(j+1,k)-a(j,k)}\qquad a(j,0)=\varepsilon^{-n}I(\varepsilon_j), a(j,-1)=0
\end{equation}
to obtain approximations to the coefficient $C_n$(in the columns with $k$ odd).

\section{Application}
The 2-loop amplitude including the numerator is processed in the following way.
As an explicit example, we consider the calculation of the 2-loop electroweak
self-energy function, $\Pi(s)$.
The function is the sum of a number of diagrams, i.e., $\Pi(s)=\sum \Pi_j(s)$ and
later we do not write the index $j$ explicitly unless it is necessary.
First, GRACE\cite{GRACE, GRACE0, GRACE1, GRACE2} 
automatically generates the diagrams.  Then  
for each diagram its numerator and denominator are  
given by a symbolic code in REDUCE
to compute the integral:
\begin{equation}
\Pi(s)=\int[d\ell_1][d\ell_2]\frac{\mathbf{N}(\ell_1,\ell_2)}{P_1 P_2\cdots P_N}
=\int dx_1dx_2\cdots dx_N\delta(1-\sum x_k) G
\end{equation}
where $\mathbf{N}$ is the numerator and  $P_k=p_k^2-m_k^2+i \rho$. Here,
\begin{equation}
G=\Gamma(N)\int[d\ell_1][d\ell_2]\frac{\mathbf{N}(\ell_1,\ell_2)}{\Delta^N}
\end{equation}
where
$\Delta = {}^{\mathrm{t}}\vec{\ell}A\vec{\ell} +2 {}^{\mathrm{t}}\vec{\ell}\,\vec{b}+C$.
Here $\vec{\ell}=\displaystyle{\left(\begin{array}{c} \ell_1 \\ \ell_2 \end{array} \right)}$.
After a sequence of the variable transformation,
we have 
\begin{equation}
\Delta=\ell_1^2+\ell_2^2 -V,\quad
\mathbf{N}=f^{00}+f^{10}\ell_1^2 +f^{01}\ell_2^2 +
f^{20}(\ell_1^2)^2 + f^{11}\ell_1^2 \ell_2^2 +f^{02}(\ell_2^2)^2 \ ,
\end{equation}
and $G$ is given by
\begin{equation}
G=\sum \frac{(-1)^{N+k+m}}{(4\pi)^n} 
\frac{\Gamma(N-n-k-m)\Gamma(n/2+k)\Gamma(n/2+m)}{\left(\Gamma(n/2)\right)^2}  
\frac{f^{km}}{U^{n/2}V^{N-n-k-m}}.
\end{equation}
where $U$ and the product of $U$ and $V$ are polynomials of $x$'s, 
\begin{equation}
U=\mathrm{det} A, \quad 
UV =-\mathrm{det} \left(
\begin{array}{cc} A & \vec{b} \\ {}^{\mathrm{t}}\vec{b} & C \end{array}
\right)\ .
\end{equation}
The derivative function $d\Pi(s)/ds$ is also computed in a similar manner.
The system generates the FORTRAN code for $G$ automatically.
However, at the present status, one must prepare a part of
REDUCE code dependent on the topology of the diagram manually.
This part needs more work for the complete automation.
This point can be automated 
since the number of the topologies in 2-loop diagrams is rather limited, so that
we can prepare all possible code for these in the system.

There is a variant of the treatment of the numerator used 
in the calculation of the 2-loop electroweak correction to muon $g-2$
in ref.\cite{Nakazawa}. 
In the future, we would also like to implement this method since
it is preferable to have another method
for the automated computation to confirm the results.

When the integral has UV divergence, we keep $\varepsilon$ finite.
Also when the denominator in the integrand vanishes in the integration region,
we keep $\rho$ finite.  For an integral with both singularities, we execute
a double extrapolation: First, we fix the value of $\varepsilon$  and
compute the integral for several values of $\rho$ to estimate the 
limit as $\rho \rightarrow 0$ using either of the methods explained in the previous section.
Then, we calculate the $C$ coefficients  of series in $\varepsilon$.
A detailed discussion on the double extrapolation is found in ref.\cite{ACAT2011, MinQM}.

As an instance, we consider the Higgs 2-point function of the 2-loop order 
in the electro-weak theory
with the non-linear gauge fixing\cite{GRACE2}.
Then GRACE generates 3082 diagrams for the function 
including counter-term diagrams.
So the system should perform well to handle
such size of computation.
Some test computations have shown that this example can be
calculated within realistic computer time.

\section{Summary}

The DCM is well developed as an important tool 
to calculate the radiative corrections
in 2-loop order when it is combined with GRACE system.
Since it is based on a fully numerical method, once it is
proved to work for the scalar integral, it can 
compute the physical amplitude without any special extension of
the computational method.

As was already discussed, 
in the large scale calculation for the higher-order correction 
it is desirable to perform the calculation automatically,
so that one can avoid possible human error.
From this view point, the DCM still needs two points to be upgraded.
One is to control the selection of the series of numerical values of 
$\{\varepsilon_j\}$ using some iterative test to find the
proper values, i.e., a variation of a machine-learning method.
Second is to introduce a system to analyze the structure of
diagrams to provide the topology-dependent part of
the numerator handler.  
In a future publication we plan to present explicit
results for 2-loop processes.


\section*{Acknowledgments}
We acknowledge the support from the National Science Foundation under 
Award Number 1126438, and
the Center for High Performance Computing and Big Data at Western 
Michigan University. This work
is further supported by Grant-in-Aid for Scientific Research (15H03668 and 17K05428) 
of JSPS, and the Large Scale
Simulation Program No.16/17-21 of KEK.

\section*{References}



\end{document}